# Spin Hall magnetoresistance sensor using $Au_xPt_{1-x}$ as the spin-orbit torque biasing layer

*Yanjun Xu, Yumeng Yang, Hang Xie and Yihong Wu*

*Department of Electrical and Computer Engineering, National University of Singapore*
*4 Engineering Drive 3, Singapore 117583*

We report on investigation of spin Hall magnetoresistance sensor based on NiFe/$Au_xPt_{1-x}$ bilayers. Compared to NiFe/Pt, the NiFe/$Au_xPt_{1-x}$ sensor exhibits a much lower power consumption (reduced by about 57%), due to 80% enhancement of spin-orbit torque efficiency of $Au_xPt_{1-x}$ at an optimum composition of $x = 0.19$ as compared to pure Pt. The enhanced spin-orbit torque efficiency allows to increase the thickness of NiFe from 1.8 nm to 2.5 nm without significantly increasing the power consumption. We show that, by increasing the NiFe thickness, we were able to improve the working field range (± 0.86 Oe), operation temperature range (150°C) and detectivity (0.71 nT/$\sqrt{Hz}$ at 1 Hz) of the sensor, which is important for practical applications.

* Author to whom correspondence should be addressed: elewuyh@nus.edu.sg

In the past few decades, a variety of magnetoresistance (MR) sensors have been developed and commercialized for diverse industrial and consumer applications,[1-6] including in the rapidly developing internet-of-things (IoT) paradigm and related technologies.[7] These include the anisotropic magnetoresistance (AMR), giant magnetoresistance (GMR) and tunnel magnetoresistance (TMR) sensors.[1,8-10] All these sensors require sophisticated transverse biasing scheme for achieving linear response to an external field.[11] Recently we have demonstrated a spin Hall magnetoresistance (SMR) sensor using spin orbit torque (SOT) induced field-like effective field as the built-in linearization mechanism.[12,13] The use of SOT biasing greatly simplifies the sensor design, which consists of only a NiFe/Pt bilayer.[11] Furthermore, since the SMR is a second order effect, it allows to drive the sensor by an ac current and detect the response using the rectification technique. The combination of all these features has led to a SMR sensor with nearly zero dc offset and negligible hysteresis, and a detectivity around 1 nT/$\sqrt{Hz}$ at 1 Hz.[14] The performance is remarkable considering its extremely simple structure. However, in order to obtain a large field-like SOT effective field, in the previous studies, the NiFe layer thickness has been optimized to be around 1.8 nm. Such a small thickness of NiFe limits both the sensor's working field (±0.35 Oe) and operation temperature range (80 ºC), which may hinder practical application of the SMR sensor. Both issues, however, could be readily resolved if we can have a spin current and SOT generator which is more efficient than Pt and at the same time has a relatively low resistivity.

Recently, several works have reported that alloying Pt with Au is an effective way to increase the SOT efficiency through enhancing the intrinsic spin Hall effect,[15,16] while the resistivity of $Au_xPt_{1-x}$ alloy is still much lower than that of $\beta$-W,[17] $\beta$-Ta,[18] Pt-Hf alloy,[19] Pt/Hf multilayers[20] and topological insulators;[21,22] the latter is important for low-power operation of the sensor. In this work, we examine the possibility of using $Au_xPt_{1-x}$ alloy to improve the working field and operation temperature range of the SMR sensors. Specifically, we fabricated

Wheatstone bridge SMR sensors consisting of NiFe(2.5)/Au$_x$Pt$_{1-x}$(3.2) bilayers (the numbers inside the parenthesis are thicknesses in nm) with the same dimensions but different Au composition, and investigated how the sensor performs against NiFe/Pt sensors. It is found that the SOT efficiency of Au$_x$Pt$_{1-x}$ is about 80% enhanced as compared to Pt at an optimum composition of $x = 0.19$, with the resistivity ($\rho_{AuPt}$) maintained at 45.77 μΩ·cm. As a result, the power consumption of NiFe(2.5)/Au$_{19}$Pt$_{81}$(3.2) sensor is reduced by about 57% as compared to that of NiFe(2.5)/Pt(3.2) without compromising the sensitivity, working field and operation temperature range. Meanwhile, the NiFe(2.5)/Au$_{19}$Pt$_{81}$(3.2) sensor also shows an improved detectivity of 0.71 nT/$\sqrt{Hz}$ at 1 Hz and much larger working field and temperature ranges, as compared with the NiFe(1.8)/Pt(2) sensor reported in our previous work.[13,14]

As shown in Fig. 1(a), the Wheatstone bridge SMR sensor comprising of four ellipsoidal NiFe(2.5)/[Au($t_{Au}$)/Pt(0.8-$t_{Au}$)]$_4$ sensing elements with a long axis length of 800 μm and an aspect ratio of 4:1 was fabricated on SiO$_2$/Si substrate using combined techniques of photolithography and liftoff. The NiFe layer was deposited first followed by the deposition of Au/Pt multilayers. The use of Au/Pt multilayer instead of AuPt alloy is merely of technical reason that we only had a Pt target in the sputtering chamber and Au pallets in the evaporation chamber (the two chambers are connected by high vacuum). The composition of AuPt is controlled by the thickness of Au ($t_{Au}$) and Pt (0.8-$t_{Au}$) with the repeating period fixed at 4. All the layers were deposited in a multi-chamber system at a base pressure below 3×10$^{-8}$ Torr and a working pressure of 3×10$^{-3}$ Torr (for sputtering) without breaking the vacuum. An in-plane field of ~500 Oe was applied along the long axis of sensing elements during the deposition to induce a uniaxial anisotropy for the NiFe layer. Fig. 1(b) shows the XPS spectra for coupon thin films with a layer structure of NiFe(2.5)/[Au($t_{Au}$)/Pt(0.8-$t_{Au}$)]$_4$, where $t_{Au}$ = 0, 0.15, 0.2, 0.25, and 0.6 nm, respectively. The Pt 4f$_{7/2}$(Pt 4f$_{5/2}$) peak appears to have a gradual shift to a lower binding energy from 71.52 eV(74.85 eV) to 71.09 eV(74.39 eV) with increasing the Au layer thickness from 0 to 0.6 nm. The shift of the binding energy for Pt in Au/Pt multilayers is attributed to the charge transfer between Pt and Au due to the formation of Au/Pt alloy.[23,24] Therefore, we can consider [Au($t_{Au}$)/Pt(0.8-$t_{Au}$)]$_4$ multilayer effectively as a Au$_x$Pt$_{1-x}$ alloy with the Au composition ($x$) approximately given by $\frac{t_{Au}}{0.8}$. It is worth noting that the choosing of smallest Au thickness at 0.15 nm is due to the limitation of the deposition system used in this study.

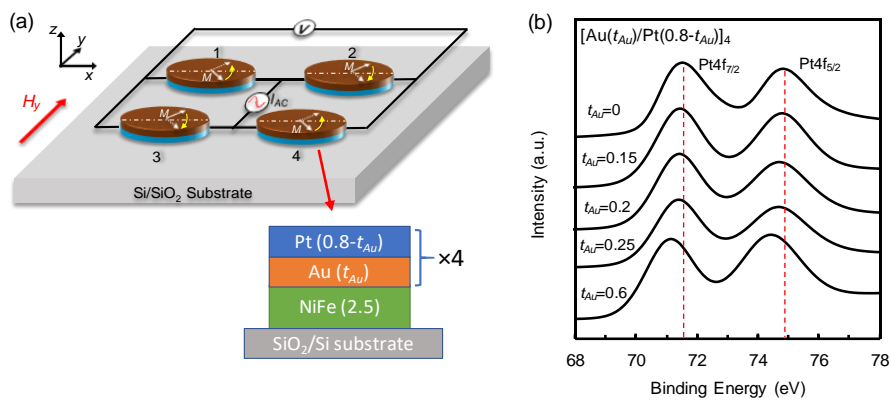

FIG.1. (a) Schematic of a Wheatstone bridge SMR sensor comprising of four ellipsoidal NiFe(2.5)/[Au($t_{Au}$)/Pt(0.8-$t_{Au}$)]$_4$ bilayer sensing elements with the arrows indicating the magnetization direction driven by the ac current. (b) XPS spectra of Pt 4f peaks for coupon thin films with a layer structure of NiFe(2.5)/[Au($t_{Au}$)/Pt(0.8-$t_{Au}$)]$_4$, where $t_{Au}$ = 0, 0.15, 0.2, 0.25, and 0.6 nm, respectively.

After confirming the Au-Pt alloying phase, we characterized the SOT effective field in NiFe(2.5)/Au$_x$Pt$_{1-x}$(3.2) bilayers. Fig. 2(a) shows the MR signal as a function of external field for one of the four NiFe(2.5)/Au$_{19}$Pt$_{81}$(3.2) sensing elements at different bias current densities in the Au$_x$Pt$_{1-x}$ layer ($j_{AuPt}$). The field is swept in $y$-direction. The general observations are: 1) there is a clear shift of the MR curve in the field direction due to the bias current; 2) the magnitude of the shift increases with increasing the bias current density; 3) the direction of shift is opposite for bias current with opposite directions. All these features point to the existence of an effective bias field ($H_{bias}$) in $y$-direction, which is the sum of the SOT induced field-like effective field ($H_{FL}$) and the Oersted field ($H_{Oe}$).[13] In the inset of Fig. 2(a), we plot the field shift ($\Delta H$) against the bias current density from which we obtain a slope of $H_{bias}/j_{AuPt}$= 0.8 Oe/(10$^6$A/cm$^2$). A simple parallel resistor model has been used in this work to calculate the current density in Au$_x$Pt$_{1-x}$ layer. The almost linear dependence on current density clearly rules out Joule heating as the cause for the shift, because otherwise the field shift would have a quadratic dependence on $I$. The Oersted field in the middle of NiFe layer is approximated by $H_{Oe} = \frac{1}{2} t_{AuPt} j_{AuPt}$, where $t_{AuPt}$ is the thickness of Au$_x$Pt$_{1-x}$. In the present case, $t_{AuPt}$= 3.2 nm, therefore $H_{Oe}/j_{AuPt}$= 0.2 Oe/(10$^6$A/cm$^2$), which is much smaller than the extracted value of $H_{bias}/j_{AuPt}$. The net SOT effective field $H_{FL}$ can be obtained from $H_{bias}$ by subtracting the Oersted field.

Fig. 2(b) shows $H_{FL}/j_{AuPt}$ for NiFe(2.5)/Au$_x$Pt$_{1-x}$(3.2) bilayers with different Au composition. As can be seen, the SOT efficiency, which is 0.34 Oe/(10$^6$A/cm$^2$) at $x = 0$ (pure Pt), increases to 0.6 Oe/(10$^6$A/cm$^2$) at $x = 0.19$ and then gradually drops to 0.002 at $x = 1$ (pure Au). The SOT efficiency for NiFe(2.5)/Au$_{0.19}$Pt$_{0.81}$(3.2) is almost 80% larger than that of NiFe(2.5)/Pt(3.2) and is comparable to the value for NiFe(1.8)/Pt(2) reported previously,[13,14] despite the much smaller $M_s t_{NiFe}$ for NiFe (1.8). Considering the fact that the SOT is originated from spin current generated in the heavy metal (HM) layer by the spin Hall effect (SHE)[25-27] and contributes to both field-like and damping-like effective field,[28-33] the SOT induced field-like effective field can be estimated as $\frac{H_{FL}}{j_{AuPt}} = \frac{\sigma_{SH} T_{int} \rho_{AuPt}}{\mu_0 M_s t_{NiFe}}$, where $\sigma_{SH}$ the spin Hall conductivity of HM, $T_{int}$ the spin transparency of the HM/FM interface, $\mu_0$ the vacuum permeability and $t_{NiFe}$ the thickness of NiFe.[16,28,31] Since $M_s$ and $t_{NiFe}$ are fixed, the nonmonotonic $x$ dependence of SOT efficiency in Fig. 2(b) can be interpreted by the change in $\rho_{AuPt}$ and apparent spin Hall conductivity ($\sigma_{SH}^* = \sigma_{SH} T_{int}$).[16,34] Fig. 2(c) shows the resistivity of 3.2 nm Au$_x$Pt$_{1-x}$ alloy extracted using the parallel resistor model with $\rho_{NiFe}$= 78.77 μΩ·cm and $\rho_{Pt}$= 31.66 μΩ·cm. It exhibits a broad positive hump with the maximum value around $x =$ 0.25 due to enhanced electron scattering at the specific alloy concentration range.[16] We then calculated the apparent spin Hall conductivity $\sigma_{SH}^*$ as a function of $x$ using the experimental results in Fig. 2(b) and 2(c) with $\mu_o M_s = 0.8$ T (experimental value) and $t_{NiFe} = 2.5\ nm$. The calculated values are shown in Fig. 2(d) which indicate that, instead of decreasing with increasing the resistivity,[35] $\sigma_{SH}^*$ increases by about 15% at $x = 0.19$ as compared with pure Pt. This result is consistent with calculated intrinsic spin Hall conductivity $\sigma_{SH}^{int}$ for bulk $Au_xPt_{1-x}$,[15] suggesting that the intrinsic spin Hall contribution is dominant in $Au_xPt_{1-x}$ alloys.

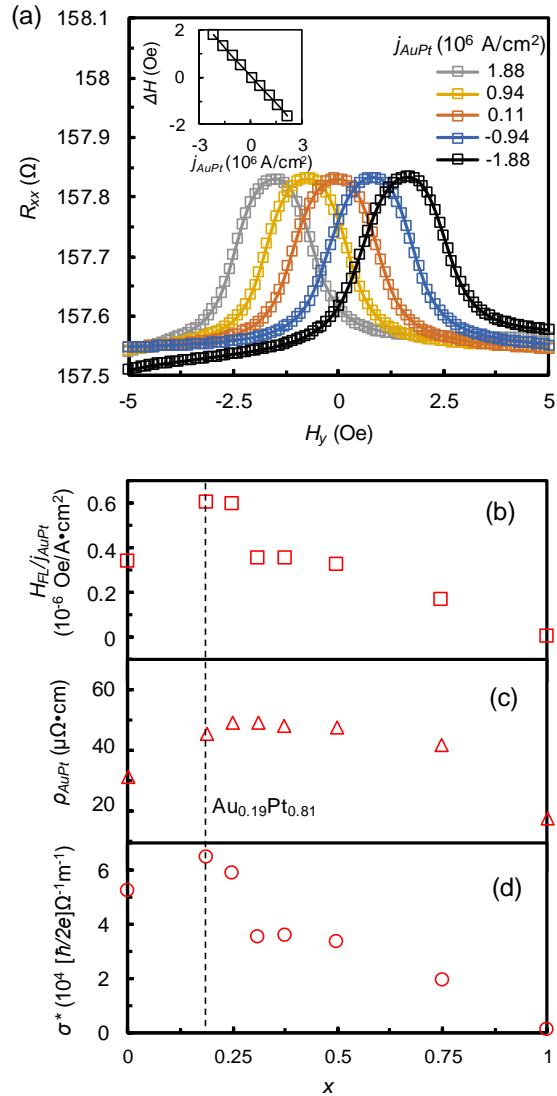

FIG.2. (a) MR response of one of the NiFe(2.5)/Au$_{19}$Pt$_{81}$(3.2) sensing elements of the Wheatstone bridge sensor measured at different current densities. Inset: Field shift as a function of bias current density. (b)-(d) Au composition dependence of SOT efficiency $H_{FL}/j_{AuPt}$(b), resistivity $\rho_{AuPt}$(c) and the apparent spin Hall conductivity $\sigma_{SH}^*$(d) for $Au_xPt_{1-x}$ alloy.

We now turn to the sensor performance of the Wheatstone bridge SMR sensor based on NiFe(2.5)/Au$_x$Pt$_{1-x}$(3.2) bilayers. As shown schematically in Fig. 1(a), the sensor is driven by an ac current and its response to an external magnetic field ($H_y$) is detected as a dc voltage. The detailed sensing mechanism and method to quantify various sensor performance parameters, including sensitivity, working field range, power consumption and detectivity, can be found in our previous work.[13,14] In order to minimize the influence of earth field, both the sensor and Helmholtz coils for generating $H_y$ were placed inside a magnetically shielded cylinder. The frequency of the ac bias current ($f_b$) applied to the sensor is fixed at 5000 Hz, unless specified. Fig. 3(a)-(c) summarizes the sensitivity, working field range and power consumption of sensors with NiFe(2.5)/Au$_x$Pt$_{1-x}$(3.2) sensing elements of same dimension ($800\ \mu m \times 200\ \mu m$ ellipsoid) but different Au composition. As shown in Fig. 3(a), the sensitivity of the sensor does not change much for $x < 0.25$, around 1.0 ~ 1.1 mV/V/Oe, but it starts to drop quickly as $x$ increases further. The overall trend corroborates well the $x$-dependence of $\sigma_{SH}^*$ shown in Fig. 2(d). This is understandable because the MR ratio of FM/HM bilayers consists of contributions from both AMR and SMR; the former does not depend much on the HM layer except for the current shunting effect, but the latter is affected strongly by both the spin Hall conductivity of the HM layer and the transparency of the FM/HM interface (*i.e.*, $\sigma_{SH}^*$). Another important performance parameter for any type of MR sensor is the working field range, which in this study is defined as the field range within which the sensor exhibits a linearity error smaller than 6%. As shown in Fig. 3(b), the working field range only varies slightly between ±0.84 and ±0.93 Oe in the entire Au composition range. In conventional Wheatstone bridge type of AMR sensors, all the sensing elements operate in the linear regime, *i.e.*, the magnetization rotates around 45º with respect to the current. In contrast, in the case of ac-driven SMR sensor, each sensing element operates in the nonlinear region, *i.e.*, the magnetization rotates around 0º with respect to the current. Nonlinearity will appear when the external field becomes comparable to the peak amplitude of the SOT effective field or the sum of the external field and SOT effective field approaches 45° with respect to the current, whichever occurs earlier. When this happens the sensitivity of the sensor will also drop. Although the linear range can be readily expanded by increasing the hardness of the FM layer, in practice, one has to strike a balance between sensitivity, working field range and power consumption. As can be seen in Fig. 3(c), the NiFe(2.5)/Au$_{0.19}$Pt$_{0.81}$(3.2) sensor exhibits the lowest power consumption. This can be explained because the power consumption ($P$) of the Au$_x$Pt$_{1-x}$ based SMR sensor is given by $P = 4j_{AuPt}^2 \rho_{AuPt} t_{AuPt} Lw(1 + \frac{\rho_{AuPt} t_{AuPt}}{\rho_{NiFe} t_{NiFe}})$, where $L$ ($w$) is the length (width) of the sensing element. Since $j_{AuPt} = \frac{\mu_o M_s t_{NiFe} H_{FL}}{\sigma_{SH} T_{int} \rho_{AuPt}}$, the power consumption can be further derived in terms of apparent spin Hall conductivity $\sigma_{SH}^*$ and resistivity of Au$_x$Pt$_{1-x}$ layer as $P = 4 \frac{(\mu_o M_s t_{NiFe} H_{FL})^2}{\sigma_{SH}^{*2} \rho_{AuPt}} t_{AuPt} Lw \left(1 + \frac{\rho_{AuPt} t_{AuPt}}{\rho_{NiFe} t_{NiFe}}\right)$. Therefore, when other parameters are fixed, we can expect a decrease of power consumption with an increase of $\rho_{AuPt}$ and $\sigma_{SH}^*$. This explains why the NiFe(2.5)/Au$_{0.19}$Pt$_{0.81}$(3.2) exhibits the lowest power consumption. To have an idea on how the AuPt alloy can enhance sensor performance as compared with Pt-based sensor, we compare the field responses of Pt and AuPt-based sensors with the same layer thickness in Fig. 3(d), *i.e.,* NiFe(2.5)/Au$_{0.19}$Pt$_{0.81}$(3.2) and NiFe(2.5)/Pt(3.2). The external field was swept in *y*-direction from -2 Oe to +2 Oe and then back to -2 Oe. It is observed that the forward and backward sweeping response curves nearly overlap with each other for both sensors, indicating a negligible hysteresis in the full field range. In addition, as shown in the inset, the dc offset is also nearly zero. The working field ranges are ±0.84 Oe and ±0.86 Oe for the NiFe(2.5)/Au$_{0.19}$Pt$_{0.81}$(3.2) and NiFe(2.5)/Pt(3.2) sensors, respectively. Within the linear range, both sensors exhibit almost the same sensitivity of around 1.10 mV/V/Oe. Despite the similar sensor performances, the root-mean-square (rms) amplitude of the ac bias current density required to obtain a linear response with maximum sensitivity is greatly reduced from $1.8\times10^6$ A/cm$^2$ for NiFe(2.5)/Pt(3.2) to $9.4\times10^5$ A/cm$^2$ for NiFe(2.5)/Au$_{0.19}$Pt$_{0.81}$(3.2). As a result, the NiFe(2.5)/Au$_{0.19}$Pt$_{0.81}$(3.2) sensor could achieve up to 57% reduced power consumption without compromising sensitivity and working field range, as compared with NiFe(2.5)/Pt(3.2) sensor. Besides, we also investigated the operation temperature range of both NiFe(2.5)/Au$_{0.19}$Pt$_{0.81}$(3.2) and NiFe(2.5)/Pt(3.2) sensors. As shown in Fig. 3(e), the temperature dependences of the sensitivities of both sensors follow similar trend: they remain quite stable up to 60 ºC, and then start to drop with the increase of temperature until finally become diminishingly small at around 150 ºC. This dependence can be readily explained by the temperature dependence of the MR of the FM layer when approaching the blocking or Curie temperature.[36] Since the sensitivity of the sensor is mainly determined by the size of the MR, it would drop as the temperature is approaching the Curie temperature of NiFe. Furthermore, it is worth noting that the operation temperature range has been significantly improved from 80 ºC for NiFe(1.8)/Pt(2) in our previous work[14] up to 150 ºC for NiFe(2.5)/Au$_{0.19}$Pt$_{0.81}$(3.2), due to the higher blocking or Curie temperature of thicker NiFe.[37]

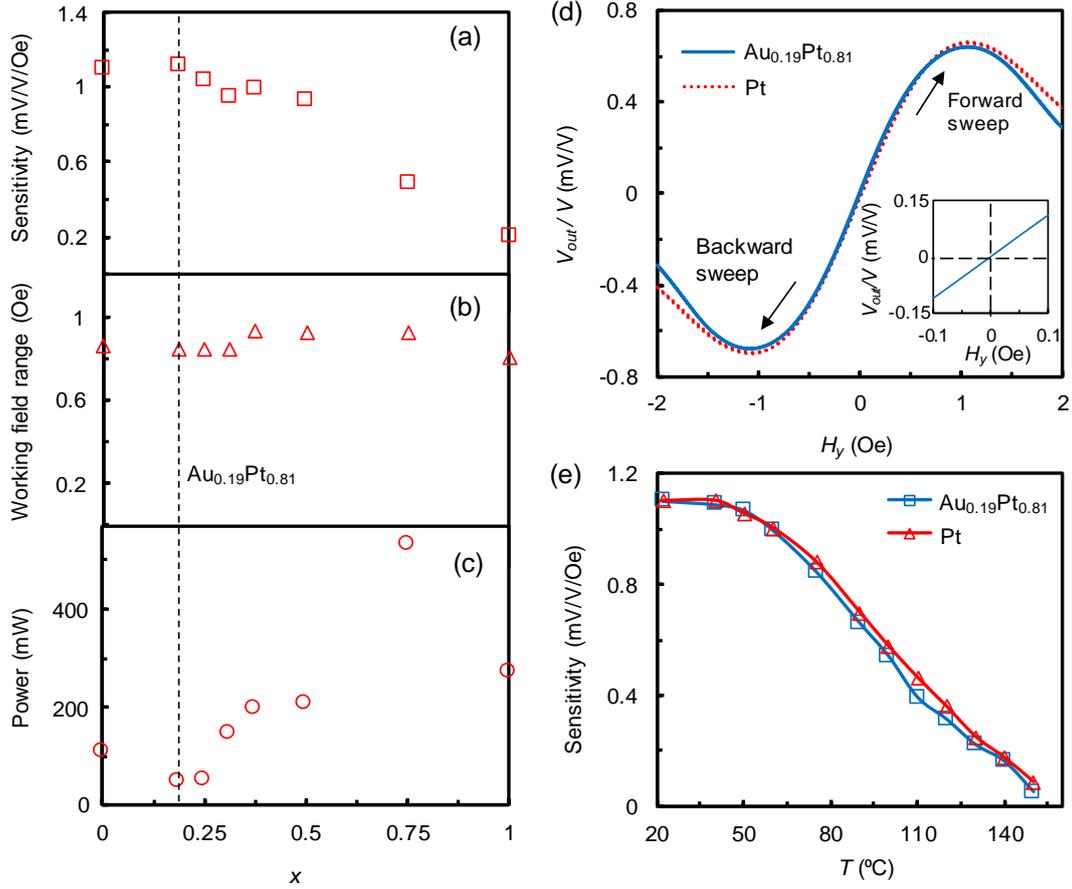

FIG.3. (a)-(c) Au composition dependence of sensitivity(a), working field range(b) and power consumption(c) for NiFe(2.5)/Au$_x$Pt$_{1-x}$(3.2) sensor. (d) Response of SMR sensor to external field in *y*-direction swept from -2 Oe to +2 Oe and then back to -2 Oe. (solid-line: NiFe(2.5)/Au$_{0.19}$Pt$_{0.81}$(3.2) and dashed-line: NiFe(2.5)/Pt(3.2)). The rms amplitude of the applied ac bias current density is $9.4\times10^5$ A/cm$^2$ for NiFe(2.5)/Au$_{0.19}$Pt$_{0.81}$(3.2) and $1.8\times10^6$ A/cm$^2$ for NiFe(2.5)/Pt(3.2). (e) Temperature dependence of sensitivity for NiFe(2.5)/Au$_{0.19}$Pt$_{0.81}$(3.2) (blue squares) and NiFe(2.5)/Pt(3.2) (red triangles) sensors.

Finally, we characterized the detectivity of the NiFe(2.5)/Au$_{0.19}$Pt$_{0.81}$(3.2) sensor. Fig. 4(a) shows the detectivity of the sensor under dc and ac bias at different frequencies. The rms current density of the ac bias was fixed at $9.4\times10^5$ A/cm$^2$ for different frequencies and is the same as the dc current density. As can be seen from Fig. 4(a), the sensor under ac bias, at all bias frequencies from 2 kHz to 50 kHz, exhibits better detectivity in the low frequency range than that under dc bias. This is because the ac excitation in SMR sensor can effectively reduce the 1/*f* magnetic noise due to diminishing hysteresis.[14] To further quantify the noise performance, we extract the detectivity of the sensor at 1 Hz as a function of bias frequency and show the results in Fig. 4(b). The dc biased sensor exhibits a detectivity of about 2.1 nT/√Hz at 1 Hz (solid triangle in Fig. 4b). In comparison, the detectivity of ac biased sensor at 1 Hz falls into the range between 0.7 and 0.9 nT/√Hz at different bias frequencies (square), while the best detectivity of 0.71 nT/√Hz at 1 Hz can be achieved at a bias frequency of 5000 Hz.

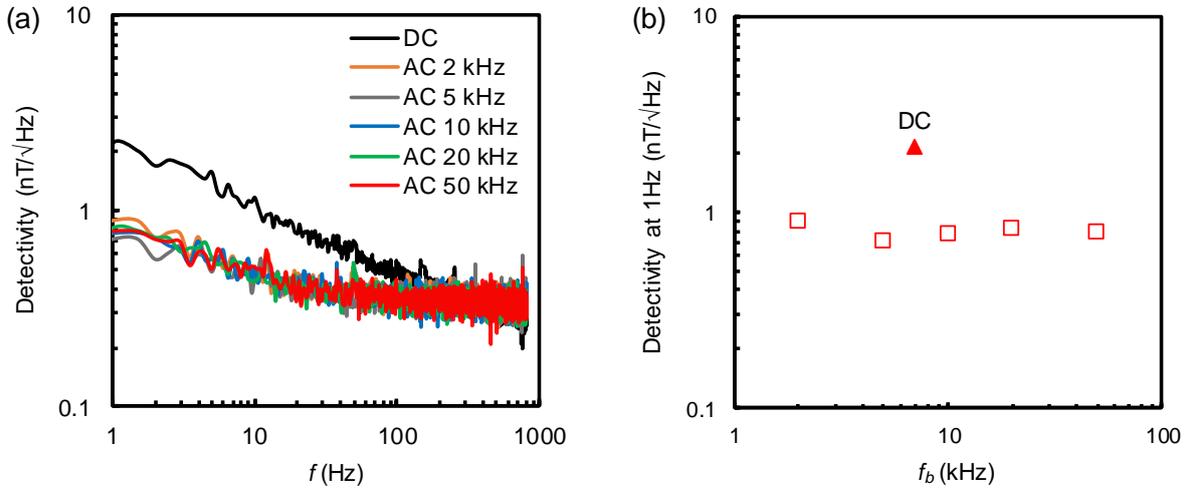

FIG.4. (a) Detectivity of the NiFe(2.5)/Au$_{0.19}$Pt$_{0.81}$(3.2) sensor under dc and ac bias at different frequencies. The rms current density of the ac bias was fixed at $9.4\times10^5$ A/cm$^2$ for different frequencies and is the same as the dc current density. (b) Extracted detectivity of the NiFe(2.5)/Au$_{0.19}$Pt$_{0.81}$(3.2) sensor at 1 Hz as a function of ac bias frequency (square). The detectivity at 1 Hz under dc bias is shown as solid triangle for comparison.

In summary, we have demonstrated that the SOT efficiency of Au$_x$Pt$_{1-x}$ alloy can be enhanced up to 80% compared with pure Pt at an optimum composition of $x = 0.19$. When being applied to Wheatstone bridge SMR sensors, it was found that at a same NiFe thickness of 2.5 nm and HM thickness of 3.2 nm (with the HM being either Pt or Au$_{19}$Pt$_{81}$), the Au$_{19}$Pt$_{81}$-based sensor exhibits 57% lower power consumption, as compared to the Pt-based sensor. Meanwhile, the working field and operation temperature range of NiFe(2.5)/Au$_{19}$Pt$_{81}$(3.2) sensor is almost twice as large as that of the NiFe(1.8)/Pt(2) sensor reported in our previous work. Furthermore, sub-nanotesla detectivity of 0.71 nT/√Hz at 1 Hz is achieved in a NiFe(2.5)/Au$_{19}$Pt$_{81}$(3.2) sensor. We believe the sensor performance can be further improved once materials with even larger SOT efficiency are identified in future. As an ending note, we want to point out that the same bridge structure can also be used to characterize SOT effective fields in FM/HM bilayer.[38-40]

**Acknowledgements**

The authors would like to acknowledge the funding support by Singapore Ministry of Education, under its AcRF Tier 2 Grant (Grant no. MOE2017-T2-2-011 and MOE2018-T2-1-076).

**Figure Captions**

FIG.1. (a) Schematic of a Wheatstone bridge SMR sensor comprising of four ellipsoidal NiFe(2.5)/[Au($t_{Au}$)/Pt(0.8-$t_{Au}$)]$_4$ bilayer sensing elements with the arrows indicating the magnetization direction driven by the ac current. (b) XPS spectra of Pt 4f peaks for coupon thin films with a layer structure of NiFe(2.5)/[Au($t_{Au}$)/Pt(0.8-$t_{Au}$)]$_4$, where $t_{Au}$ = 0, 0.15, 0.2, 0.25, and 0.6 nm, respectively.

FIG.2. (a) MR response of one of the NiFe(2.5)/Au$_{19}$Pt$_{81}$(3.2) sensing elements of the Wheatstone bridge sensor measured at different current densities. Inset: Field shift as a function of bias current density. (b)-(d) Au composition dependence of SOT efficiency $H_{FL}/j_{AuPt}$(b), resistivity $\rho_{AuPt}$(c) and the apparent spin Hall conductivity $\sigma_{SH}^*$(d) for Au$_x$Pt$_{1-x}$ alloy.

FIG.3. (a)-(c) Au composition dependence of sensitivity(a), working field range(b) and power consumption(c) for NiFe(2.5)/Au$_x$Pt$_{1-x}$(3.2) sensor. (d) Response of SMR sensor to external field in *y*-direction swept from -2 Oe to +2 Oe and then back to -2 Oe. (solid-line: NiFe(2.5)/Au$_{0.19}$Pt$_{0.81}$(3.2) and dashed-line: NiFe(2.5)/Pt(3.2)). The rms amplitude of the applied ac bias current density is 9.4×10$^5$ A/cm$^2$ for NiFe(2.5)/Au$_{0.19}$Pt$_{0.81}$(3.2) and 1.8×10$^6$ A/cm$^2$ for NiFe(2.5)/Pt(3.2). (e) Temperature dependence of sensitivity for NiFe(2.5)/Au$_{0.19}$Pt$_{0.81}$(3.2) (blue squares) and NiFe(2.5)/Pt(3.2) (red triangles) sensors.

FIG.4. (a) Detectivity of the NiFe(2.5)/Au$_{0.19}$Pt$_{0.81}$(3.2) sensor under dc and ac bias at different frequencies. The rms current density of the ac bias was fixed at 9.4×10$^5$ A/cm$^2$ for different frequencies and is the same as the dc current density. (b) Extracted detectivity of the NiFe(2.5)/Au$_{0.19}$Pt$_{0.81}$(3.2) sensor at 1 Hz as a function of ac bias frequency (square). The detectivity at 1 Hz under dc bias is shown as solid triangle for comparison.